\documentclass[sigconf]{acmart}

\usepackage{graphicx}
\usepackage{subcaption}
\usepackage{multirow}
\usepackage{amsmath,booktabs}
\usepackage{pgfplots}
\pgfplotsset{compat=1.18}
\AtBeginDocument{%
  }

\copyrightyear{2026}
\acmYear{2026}
\setcopyright{cc}
\setcctype{by}
\acmConference[UbiComp Companion '26]{Companion of the 2026 ACM International Joint Conference on Pervasive and Ubiquitous Computing}{October 11--15, 2026}{Shanghai, China}
\acmBooktitle{Companion of the 2026 ACM International Joint Conference on Pervasive and Ubiquitous Computing (UbiComp Companion '26), October 11--15, 2026, Shanghai, China}
\acmDOI{10.1145/3798063.3837198}
\acmISBN{979-8-4007-2533-3/2026/10}
\begin{document}

\title{PromptShield Home: Ambient Multimodal Prompt Injection Defense for Smart-Home Agents}

\author{He Zhang}
\email{hpz5211@psu.edu}
\orcid{0000-0002-8169-1653}
\affiliation{%
\department{College of Information Sciences and Technology}
  \institution{The Pennsylvania State University}
  \city{University Park}
  \state{Pennsylvania}
  \country{USA}
  \postcode{16801}
}

\author{Feilong Li}
\email{lifeilong@chd.edu.cn}
\orcid{0009-0002-4996-0339}
\affiliation{%
\department{School of Transportation Engineering}
  \institution{Chang'an University}
  \city{Xi'an}
  \state{Shaanxi}
  \country{China}
  \postcode{710018}
}

\author{Dingning Long}
\email{dingningll@outllook.com}
\orcid{0009-0003-1495-9626}
\affiliation{%
\department{Faculty of Psychology}
  \institution{Beijing Normal University}
  \city{Beijing}
  \country{China}
  \postcode{100875}
}

\author{Yilin Cui}
\email{yilincu2@andrew.cmu.edu}
\orcid{0009-0003-2063-3411}
\affiliation{%
\department{Heinz College of Information Systems and Public Policy}
  \institution{Carnegie Mellon University}
  \city{Pittsburgh}
  \state{Pennsylvania}
  \country{USA}
  \postcode{15213}
}

\author{Peijun Zhang}
\email{ah230zang@163.com}
\orcid{0009-0006-2547-4896}
\affiliation{%
\department{The Institute for Design Informatics}
  \institution{University of Edinburgh}
  \city{Edinburgh}
  \country{United Kingdom}
  \postcode{EH8 9BT}
}

\author{Yuewen Zhang}
\email{yw-zhang24@mails.tsinghua.edu.cn}
\orcid{0009-0007-6308-6242}
\affiliation{%
\department{The Future Laboratory}
  \institution{Tsinghua University}
  \city{Beijing}
  \country{China}
}

\author{Qianyao Xu}
\email{xuqy@mail.tsinghua.edu.cn}
\orcid{0009-0006-6707-5236}
\affiliation{%
  \institution{Tsinghua University}
  \city{Beijing}
  \country{China}
}


\author{Xinyi Fu}
\email{fuxy@tsinghua.edu.cn}
\orcid{0000-0001-6927-0111}
\authornote{Corresponding author.}
\affiliation{%
\department{The Future Laboratory}
  \institution{Tsinghua University}
  \city{Beijing}
  \country{China}
}

\renewcommand{\shortauthors}{Zhang et al.}

\begin{abstract}
Smart-home assistants increasingly use multimodal large language models (MLLMs) that perceive video and audio directly. This raises a safety question specific to the home: can the agent tell a genuine user command from ambient or externally-sourced content, television speech, on-screen text, or an overheard conversation, that merely \emph{looks} like a command? We introduce PromptShield-Home, a pilot benchmark of realistic smart-home scenarios spanning addressee ambiguity, screen/audio injection, health-monitor false triggers, mixed occupancy, and a legitimate-command floor, and use it to compare three abstraction layers: traditional detectors (L0), a single MLLM agent (L1; vision, vision$+$ASR, and audio-visual), and multi-agent mediation (L2; voting, role specialists, cross-model arbitration). Because the label distribution is skewed toward inaction, aggregate accuracy is misleading, a constant \emph{always-block} predictor scores 82\%, so we report unsafe-execution and safe-completion rates separately. The two paradigms fail in opposite ways: detectors act on everything, while every MLLM configuration \emph{over-refuses}, completing almost no genuine command and missing a true fall in every case. Crucially, their correct sets are disjoint: an \emph{oracle} that always picks the right layer reaches 94.1\%, against 76.5\% for the best single layer. We report this as an upper bound, not a system---no router is implemented---and argue that home-agent safety is best served by learned routing and sensor fusion, not by replacing detectors with an MLLM.
\end{abstract}




\maketitle
\vspace{-0.6em}
\section{Introduction}
Smart homes increasingly support safety, comfort, autonomy, and everyday convenience through connected devices, sensors, and automated control~\cite{zavei2012exploring,hanes2017iot,chan2008review,fu2023review}. Yet many deployed systems still rely on rule-based automation and Trigger-Action Programming (TAP), which become difficult to configure and maintain when domestic situations involve multiple devices, changing contexts, and complex dependencies~\cite{10.1145/3526114.3558776,li2021motivations,wilson2015smart,10.1145/3613904.3642866}.

Recent LLMs and MLLMs offer a more flexible interface: they can interpret natural language, process visual and audio evidence, reason over context, and map high-level goals to device actions~\cite{10386743,10729865,10599909,10.1145/3742414.3794710}. However, this flexibility creates a safety problem specific to the home. Domestic environments contain many ambient signals that are not meant as commands, including television speech, conversations, screen text, notes, and other visible or audible content~\cite{schoenherr2022accidental}. An MLLM may correctly perceive these signals but wrongly treat them as user intent, causing unwanted actions or privacy leakage~\cite{greshake2023indirect,liu2023prompt,liu2024formalizing,clusmann2025vlmpromptinjection,bagdasaryan2023abusingimagessoundsindirect}.

We call this threat \textit{Ambient Multimodal Prompt Injection}: visual, audio, textual, or cross-modal information in the physical environment is mistaken for an executable user command. Unlike direct prompt injection, the signal is not necessarily typed by an attacker into the model; it may naturally appear in the surrounding environment~\cite{zou2026poisononceexploitforever,shi2025lessonsdefendinggeminiindirect,xiang2026architectingsecureaiagents,307726}. The challenge is therefore not only perception, but source attribution and action authorization: the agent must decide whether observed content reflects real user intent while avoiding over-caution that ignores valid commands or emergencies~\cite{shi2024judgedeceiver}.

To study this problem, we present \textit{PromptShield-Home}, an early-stage safety-mediation framework and pilot benchmark for smart-home agents. Each case is modeled as a ternary, risk-aware decision: \emph{execute}, \emph{block}, or \emph{ask the user}. We compare three abstraction layers: traditional detectors (wake-word, pose, and OCR), a single MLLM agent under vision, vision$+$ASR, and native audio-visual settings, and multi-agent mediation via voting, role specialists, and cross-model arbitration. Our pilot scenarios cover addressee ambiguity, screen/audio injection, health false triggers, mixed occupancy, legitimate commands, and occupancy-state queries.

Because most smart-home situations should not trigger an action, aggregate accuracy is misleading: a constant ``never act'' predictor already reaches 82\% accuracy. We therefore evaluate safety and utility separately. Our results show that traditional detectors over-execute, whereas MLLMs systematically over-refuse, completing few genuine commands and missing a true fall in every tested configuration. Yet their correct decisions are complementary: an \emph{oracle} over L0 and the vision MLLM reaches 94.1\% against 76.5\% for the best single layer---an upper bound, since no router is implemented. This suggests safer home agents should combine routing and sensor fusion rather than replacing detectors with MLLMs.

\vspace{-1.2em}
\section{Methods}

\textbf{Problem.} A smart-home agent must decide, from multimodal scene
evidence and a candidate device action, whether to act, without mistaking
ambient or externally-sourced content (TV speech, on-screen text,
interpersonal talk) for a genuine user command. Each instance is
$s_i=(E_i,A_i,z_i)$ with evidence $E_i=(V_i,U_i,T_i)$ (video, audio, optional
one-line description), candidate action $A_i=(d_i,a_i)$, and hidden binary gold
$z_i\in\{0,1\}$. Every layer emits
\vspace{-0.4em}
\[
\hat{y}_i\in\mathcal{Y}=\{\texttt{execute},\,\texttt{block},\,\texttt{ask\_user}\},
\qquad
e_i=\mathbb{1}[\hat{y}_i=\texttt{execute}],
\vspace{-0.4em}\]
i.e.\ \texttt{block} and \texttt{ask\_user} both count as non-execution; invalid
output falls back to \texttt{ask\_user} (never executes).

\textbf{Three-layer comparative framework.} All layers share the same input, decision space, and metrics, so safety and utility are directly comparable. The layers are an engineering decomposition ordered by delegated semantic authority, not an existing taxonomy; intermediate designs such as learned addressee- or activity-recognition models lack action-level semantics, so we expect their errors on the L0 side and treat them as features for the router of Sec.~\ref{sec:route} rather than as a fourth layer.

\begin{center}\footnotesize
\captionof{table}{The three abstraction layers compared in PromptShield-Home.}
\label{tab:layers}
\vspace{-0.8em}
\begin{tabular}{@{}llr@{}}
\toprule
\textbf{Layer} & \textbf{Mechanism} & \textbf{Variants} \\
\midrule
L0 & traditional detectors  & wake-word / pose / OCR \\
   & (deterministic firing) & (single baseline) \\[2pt]
L1 & single MLLM  & vision (Qwen3-VL-30B) \,/ \\
   & (zero-shot, JSON) & vision+ASR (Whisper) \,/ \\
   &                    & audio-visual (Omni-7B) \\[2pt]
L2 & multi-agent  & voting (3$\times$ sampled) \,/ role specialists \\
   & (compose calls) & \,/ cross-model arbitration \\
\bottomrule
\end{tabular}
\end{center}

Prompts are built by a fixed template $P_i=\tau(T_i,W_i,A_i)$ over three
conditions that differ only in which optional fields are filled
(\textit{minimal} / \textit{+scenario} / \textit{+transcript},
where $W_i=\mathrm{ASR}(U_i)$). The L2 specialists query distinct roles
(\textit{addressee}, \textit{safety}, \textit{authorization}); arbitration uses
a text judge over the vision and audio-visual decisions.

\textbf{Metrics.} Labels are skewed to no-action, so a constant
\textit{always-block} predictor reaches high accuracy (and matches our strongest
config). We therefore report the two error modes a deployable agent must
\emph{jointly} minimize:
\[
\mathrm{UER}=\frac{|\{i:z_i{=}0\wedge e_i{=}1\}|}{|\{i:z_i{=}0\}|},
\qquad
\mathrm{SCR}=\frac{|\{i:z_i{=}1\wedge e_i{=}1\}|}{|\{i:z_i{=}1\}|},
\]
i.e.\ the \emph{unsafe-execution rate} ($\downarrow$; share of no-action cases executed) and \emph{safe-completion rate} ($\uparrow$; share of action-required cases executed). We also report the \emph{false-block rate} $\mathrm{FBR}=|\{i:z_i{=}1\wedge\hat{y}_i{=}\texttt{block}\}|/|\{i:z_i{=}1\}|$ and \emph{human-confirmation rate} $\mathrm{HCR}=|\{i:\hat{y}_i{=}\texttt{ask\_user}\}|/N$; HCR is over all $N$ cases, FBR only over action-required ones.

\begin{figure}[t]
\centering
\includegraphics[width=0.88\linewidth]{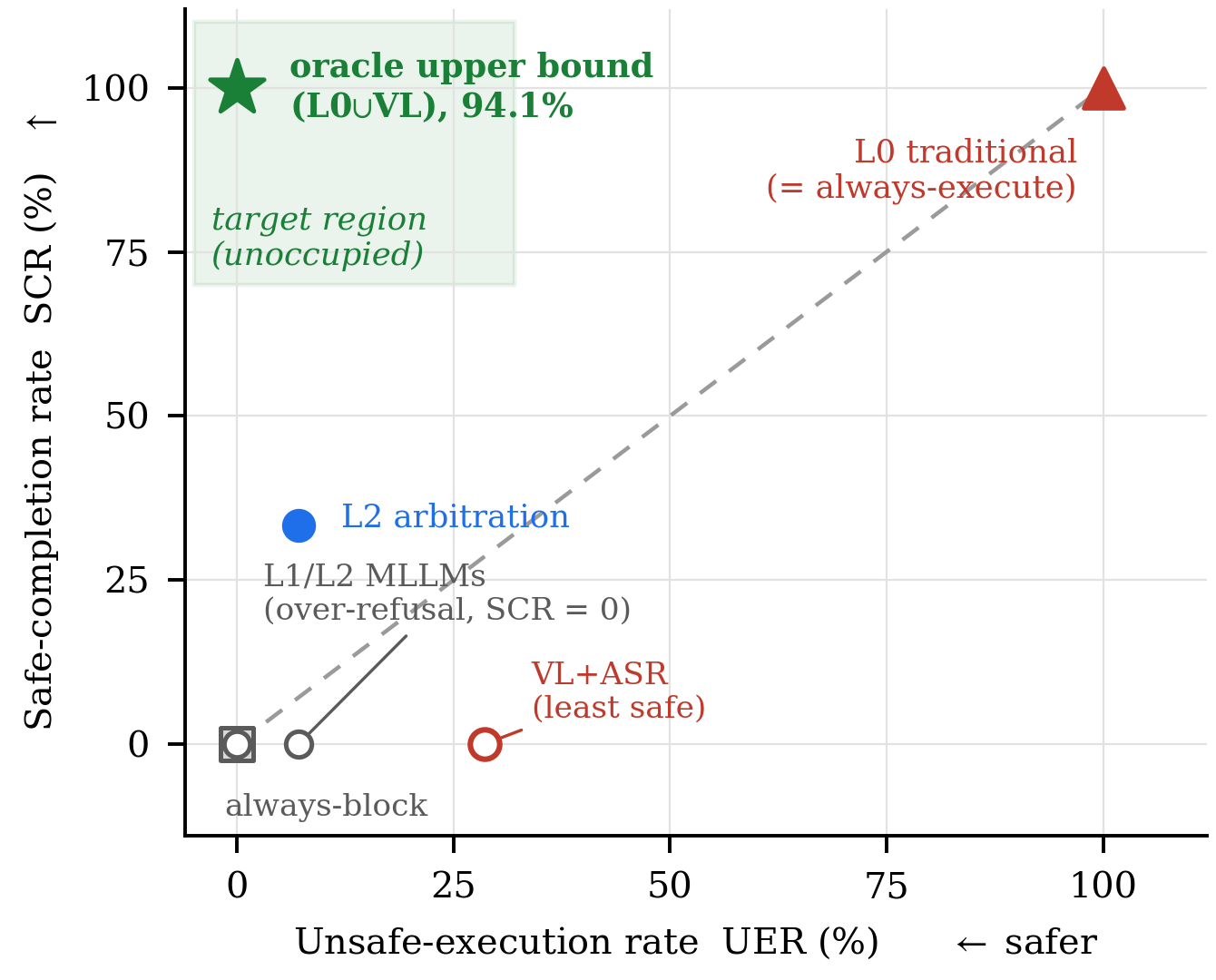}
\vspace{-1.2em}
\caption{\textbf{The safety--utility plane.} Every single- and multi-agent configuration lies on the reckless$\leftrightarrow$over-refusal diagonal: traditional detectors (L0) act on everything (top-right), while all MLLM configurations over-refuse (bottom-left, SCR\,$=$\,0; VL+ASR is the least safe). The deployable top-left corner is empty; it is reached only by an \emph{oracle} over the complementary L0 and vision-MLLM layers (94.1\%), an upper bound motivating learned routing rather than MLLM replacement.}
\label{fig:safety-utility}
\vspace{-1.2em}
\end{figure}

\begin{figure}[t]
\centering
\includegraphics[width=0.82\linewidth]{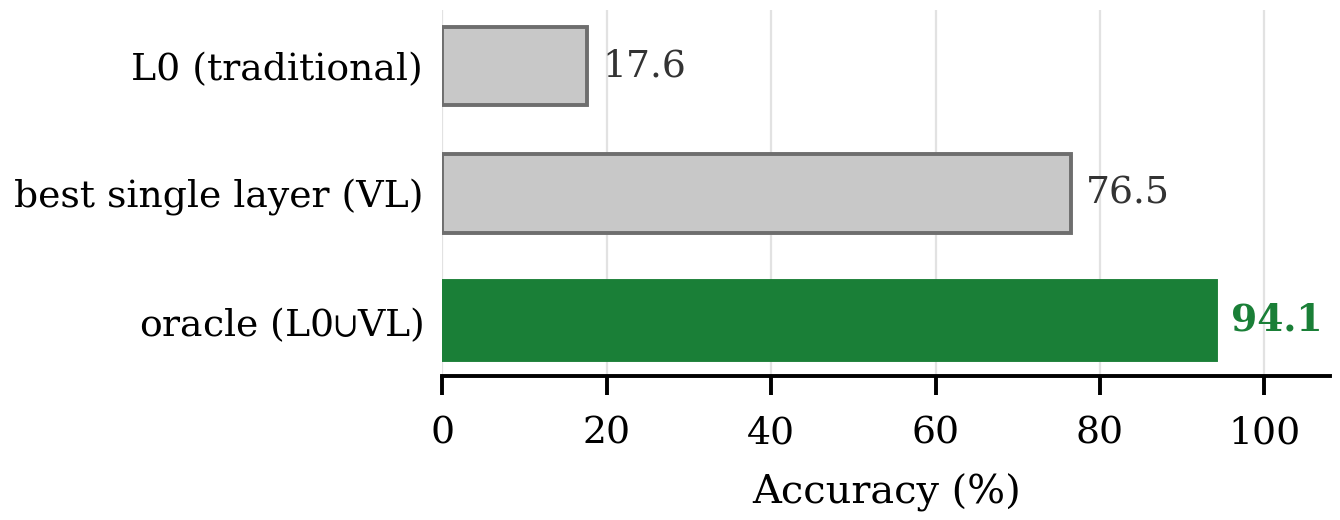}
\vspace{-1em}
\caption{\textbf{Complementarity.} L0 and the vision MLLM have disjoint correct sets, so an \emph{oracle} over the two upper-bounds at 94.1\% versus 76.5\% for the best single layer.}
\label{fig:complementarity}
\vspace{-1.6em}
\end{figure}

\vspace{-0.8em}
\section{Dataset}
\begin{figure}[t]
\centering
\includegraphics[width=0.7\linewidth]{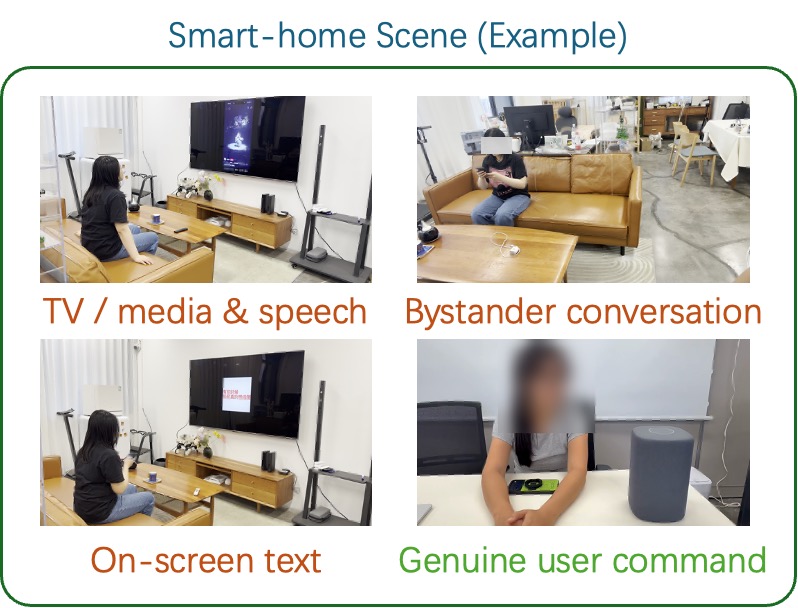}
\vspace{-0.6em}
\caption{Example smart-home scenarios in the PromptShield-Home benchmark. The scenarios cover common sources of ambient multimodal content that may be misinterpreted as executable user intent, including TV/media speech, bystander conversation, on-screen text, and genuine user commands. Faces are blurred for privacy.}
\label{fig:datasample}
\vspace{-1.6em}
\end{figure}
We designed a pilot benchmark to test whether smart-home agents can separate real user intent from ambient information that should not be executed. Guided by known smart-home failures and user risk cases, three researchers scripted each scenario in two meetings before recording, fixing the home setup, the candidate device--action pair, the intended gold decision, and the recording criteria. Two researchers then acted out and recorded the scenarios in a residential environment with a smart speaker, smart lights, home cameras, and a television, capturing visual, audio, textual, and cross-modal cues. Each clip was checked against the script and re-recorded when needed, and every clip and its gold decision were reviewed by at least five authors, entering the benchmark only on unanimous agreement; gold labels are thus fixed by construction rather than judged post hoc, which we report as a structured consensus protocol rather than an inter-annotator agreement statistic. Several scenarios are \emph{benign/harmful pairs} differing only in the key safety distinction, a real fall versus a yoga pose, an empty versus an occupied room, self-talk versus a real command, which isolates the decision the agent must make.

The benchmark contains \textbf{19 scenarios}: 17 video clips, recorded as \texttt{.MOV} or \texttt{.mp4}, and 2 still images for occupancy queries. The scenarios cover six categories, as shown in Table~\ref{tab:dataset}. Each scenario is annotated with three types of information: (i) multimodal evidence, including video, audio, and a one-line scene description; (ii) a candidate device--action pair, such as \texttt{light/turn\_off}, \texttt{emergency\_\allowbreak alert/\allowbreak trigger\_alert}, or \texttt{phone/call\_\allowbreak contact}; and (iii) a gold decision, which is hidden from the model during evaluation. Most gold decisions are no-action cases: $13$ are \texttt{block} and $1$ is \texttt{ask\_user\_\allowbreak or\_\allowbreak block}. Only $3$ scenarios require \texttt{execute}, and $2$ are occupancy-state answers. This reflects a common smart-home setting and motivates our separate evaluation of safety and utility. The 17 clips are the ternary action decisions scored below ($N{=}17$: 14 no-action, 3 action-required); the 2 occupancy images are perception queries with no candidate action.

\begin{table}[t]
\centering
\footnotesize
\caption{Composition of the PromptShield-Home pilot benchmark
(19 scenarios). ``Gold'' gives the intended decision(s).}
\label{tab:dataset}
\vspace{-0.6em}
\resizebox{\columnwidth}{!}{%
\begin{tabular}{@{}llcl@{}}
\toprule
\textbf{Category} & \textbf{Stress / risk} & \textbf{\#} & \textbf{Gold} \\
\midrule
Addressee ambiguity & speech not addressed to agent & 7 & block \\
Screen/audio injection & TV / on-screen content as command & 3 & block \\
Health false trigger & fall, pose, heart-rate, gesture & 4 & 1 execute / 3 block \\
Mixed occupancy & conflicting occupants & 1 & \texttt{ask\_user} / block \\
Command floor & genuine, clearly addressed command & 2 & execute \\
Occupancy state & static empty / occupied image & 2 & empty / occupied \\
\midrule
\textbf{Total} & & \textbf{19} & \\
\bottomrule
\end{tabular}%
}
\vspace{-1.6em}
\end{table}

\vspace{-0.6em}

\paragraph{Traditional-detector annotation.}
For each scenario, we also record what a conventional detector would do. Examples include \texttt{execute\_\allowbreak via\_\allowbreak keyword} when a wake word appears in human-to-human speech, \texttt{execute\_\allowbreak via\_pose\_\allowbreak threshold} when a yoga pose is mistaken for a fall, and \texttt{execute\_via\allowbreak\_ocr\_\allowbreak keyword} when an on-screen phrase is treated as a command. This annotation has two purposes. First, it shows how current rule-based systems can fail in these cases. Second, it provides the L0 traditional-detector baseline for comparison with the MLLM-based layers, without requiring us to run a separate detector stack. The proxy is idealised in both directions, a fielded detector has misses and false alarms of its own, so the complementarity above is an upper bound on L0's contribution too.

\vspace{-0.4em}
\section{Results}
%

\begin{table}[t]
\centering\footnotesize
\caption{Safety vs.\ utility per configuration (Acc = accuracy; other metrics as defined in Sec.~2). $n{=}17$: UER is over the 14 no-action cases, SCR and FBR over the 3 action-required cases, HCR and Acc over all 17, so these columns move in steps of 7.1, 33.3 and 5.9 points and should be read as counts. The \emph{always-block} predictor already reaches 82.4\%, motivating the UER/SCR split; L0 fires on all 17 cases and so coincides with \emph{always-execute}, since every scenario contains a surface cue a rule latches onto.}
\vspace{-0.6em}
\label{tab:main}
\begin{tabular}{@{}lrrrrr@{}}
\toprule
\textbf{Configuration} & \textbf{UER} & \textbf{SCR} & \textbf{FBR} & \textbf{HCR} & \textbf{Acc} \\
\midrule
\textit{always-execute} & 100.0 & 100.0 & 0.0 & 0.0 & 17.6 \\
\textit{always-block}   & 0.0 & 0.0 & 100.0 & 0.0 & 82.4 \\
\midrule
L0 traditional          & 100.0 & 100.0 & 0.0 & 0.0 & 17.6 \\
L1a VL, minimal         & 0.0 & 0.0 & 100.0 & 5.9 & 76.5 \\
L1a VL, +scenario       & 7.1 & 0.0 & 100.0 & 5.9 & 70.6 \\
L1b VL+ASR              & 28.6 & 0.0 & 100.0 & 0.0 & 58.8 \\
L1c Omni, minimal       & 7.1 & 0.0 & 0.0 & 94.1 & 5.9 \\
L1c Omni, +scenario     & 7.1 & 0.0 & 0.0 & 82.4 & 17.6 \\
L1c Omni, +transcript   & 0.0 & 0.0 & 0.0 & 100.0 & 5.9 \\
L2 voting               & 7.1 & 0.0 & 100.0 & 5.9 & 70.6 \\
L2 specialist           & 0.0 & 0.0 & 100.0 & 0.0 & 82.4 \\
\textbf{L2 arbitration} & 7.1 & \textbf{33.3} & 66.7 & 17.6 & 64.7 \\
\bottomrule
\end{tabular}
\vspace{-2em}
\end{table}

\textbf{Finding 1 — opposite failure modes.} L0 acts on everything
(UER\,$=$\,14/14, SCR\,$=$\,3/3); every MLLM collapses to the opposite corner
(UER\,$\le$\,1/14, SCR\,$=$\,0/3): safe but systematically \emph{over-refusing}.
No configuration is low on both axes.

\textbf{Finding 2 — accuracy is degenerate.} The top-accuracy method
(L2 specialist, 82.4\%) is byte-identical to \textit{always-block}: it outputs
\texttt{block} on all 17 inputs (51/51 agent votes), including the true fall.
High accuracy certifies nothing about utility.

\textbf{Finding 3 — modality effects.} ASR transcripts are \emph{harmful}
(L1b is the least-safe MLLM, UER\,$=$\,4/14, driven by noisy transcripts);
native audio (Omni) merely shifts mass onto \texttt{ask\_user}
(HCR up to 100\%) without improving decisions.


\begin{table}[t]
\centering\footnotesize
\caption{Complementarity of L0 and the vision MLLM. Their correct sets are disjoint, so an \emph{oracle} over \{L0,\,VL\} upper-bounds at 94.1\% accuracy versus 76.5\% for the best single layer; no router is implemented.}
\label{tab:complementarity}
\vspace{-1.2em}
\begin{tabular}{@{}lcc@{}}
\toprule
& \textbf{correct set} & \textbf{count} \\
\midrule
L0 only        & V11, V16, V17 & 3 \\
VL only        & V01--V10\,$\setminus$\,V06,\,V12--V15 & 13 \\
both           & --- & 0 \\
\midrule
\textbf{oracle (L0$\cup$VL)} & & \textbf{94.1\%} \\
best single layer (VL)             & & 76.5\% \\
\bottomrule
\end{tabular}
\vspace{-1.2em}
\end{table}
\vspace{-0.2em}

\textbf{Complementarity: route, don't replace.}
L0 and the vision MLLM have \emph{disjoint} correct sets: L0 covers exactly the three action-required cases the MLLM over-refuses, the MLLM the semantic (addressee / injection) cases detectors cannot reason about, and no case is solved by both. An \emph{oracle} over \{L0, VL\} therefore upper-bounds at \textbf{94.1\%} against 76.5\% for the best single layer; no router is implemented (Sec.~\ref{sec:route}). The most safety-critical error, missing the real fall (V11), is made by \emph{every} MLLM (9/9) and caught only by the physical-sensor baseline, arguing for sensor$+$MLLM fusion (high-sensitivity trigger $+$ MLLM false-alarm filter) rather than a camera-only agent.

\textbf{Frame budget does not fix it (and costs latency).} Sweeping frame count $\times$ per-frame resolution: more frames do \emph{not} recover the missed actions (V11, V16 stay blocked at all budgets) and \emph{regress} a spatial injection case (V09 flips to unsafe \texttt{execute} at $\geq$64 frames as per-frame detail drops), while total latency grows $4.9\!\to\!8.6$\,s ($16\!\to\!128$ frames). Token count is misleading here ($409\!\to\!867$, capped budget); latency is the fair cost. The optimal frame budget is thus scenario-dependent, and denser sampling is not a remedy for over-refusal~\cite{10.1145/3746270.3760233}.

\textbf{Robustness (5 seeds).} Over-refusal is a \emph{stable} bias, not noise: the vision configurations barely vary, whereas Omni flips 16/17 decisions across seeds, so its apparent ``caution'' is instability, not principled deferral.

\begin{table}[t]
\centering\footnotesize
\caption{Robustness across 5 sampling seeds (mean\,$\pm$\,std). Vision
configurations are stable; Omni varies widely (16/17 scenarios flip decision
across seeds), indicating its high \texttt{ask\_user} rate is decision
instability rather than principled caution.}
\label{tab:robustness}
\vspace{-1.4em}
\begin{tabular}{@{}lccc@{}}
\toprule
\textbf{Config (5 seeds)} & \textbf{UER} & \textbf{SCR} & \textbf{FBR} \\
\midrule
VL minimal       & $0.0\pm0.0$  & $0.0\pm0.0$ & $86.7\pm16.3$ \\
VL +scenario     & $7.1\pm0.0$  & $0.0\pm0.0$ & $100.0\pm0.0$ \\
Omni +scenario   & $12.9\pm5.3$ & $0.0\pm0.0$ & $40.0\pm24.9$ \\
\bottomrule
\end{tabular}
\vspace{-1.6em}
\end{table}

\vspace{-0.8em}

\section{Discussion}

\subsection{Single-layer methods struggle to balance safety and utility.}
Neither paradigm alone balances the two axes. Detectors handle all action-required cases but act on almost everything (UER 14/14), whereas MLLM configurations drive unsafe execution to near zero only by refusing almost all genuine commands (SCR 0/3). The system shifts from over-execution to systematic over-refusal, and in a home both an unintended door unlock and a missed fall are unacceptable.

\vspace{-0.8em}
\subsection{Route rather than replace.}\label{sec:route}
Traditional detectors and MLLMs are complementary rather than competing: their correct decisions are disjoint, and an oracle over the two layers reaches $94.1\%$ accuracy versus $76.5\%$ for the best single layer (the only configuration above that, $82.4\%$, is the degenerate \textit{always-block} equivalent of Finding 2). Smart-home agents should therefore route each decision to the layer best suited for it rather than replace one with the other, e.g., using a high-sensitivity physical sensor (IMU, accelerometer, or radar) for fall \emph{onset detection} and an MLLM as a high-specificity \emph{false-alarm filter}~\cite{wang2020elderlyfall,10.1145/3629606.3629638}. Rule- and role-based policies remain useful as partial safeguards, but cannot anticipate all home contexts, so LLMs should act as constrained semantic mediators under explicit routing, risk, and source-attribution policies~\cite{weng2026argus}.

\vspace{-0.8em}
\subsection{More modalities do not always help.}
Adding ASR made the vision model less safe (UER 4/14) by introducing noisy transcripts, while native audio shifted decisions to \texttt{ask\_user} without resolving the case, instability rather than caution, since 16 of 17 scenarios changed across seeds whereas the vision-only over-refusal was stable. The bottleneck is calibration and source attribution, not the number of modalities.

\vspace{-0.8em}
\subsection{Conservative behavior is not the same as intelligent behavior.}
MLLM agents can reduce unsafe execution by refusing to act, but this does not make them more useful or intelligent, since they fail valid commands and even miss emergencies. Overall accuracy hides this: our highest-accuracy configuration behaves like an \textit{always-block} predictor. Safety-critical evaluation should therefore report safety and utility separately, and deployable agents must know when to act, refuse, or ask the user, as in PromptShield-Home's layered design.
\vspace{-0.8em}
\section{Limitations and Future Work}
This pilot is limited by its small benchmark ($N=19$), uneven category coverage, gold decisions verified by author consensus rather than independent annotation, and mostly single-run evaluation. The complementarity gain relies on an \emph{oracle} rather than a trained router, and the sensor-fusion argument on the L0 proxy rather than measured IMU or radar streams. Inference used one A100-80GB; the 30B vision model needs ${\sim}60$\,GB of VRAM and cannot run always-on in a home, which is itself an argument for routing. Future work will scale the benchmark, add independent annotations, train calibrated routers, and evaluate on-device sensor fusion.
\vspace{-1em}
\section{Conclusion}

PromptShield-Home provides an early benchmark for ambient multimodal prompt injection in smart-home agents. Traditional detectors over-execute, whereas MLLMs over-refuse, making aggregate accuracy insufficient. These complementary failure modes motivate source-aware routing and sensor-fusion designs rather than replacing detectors with MLLMs.



\section*{GenAI Usage Disclosure}
We used generative AI tools for language polishing and grammar correction. Generative AI models were also used as part of the experimental pipeline because this study evaluates MLLM-based smart-home agents and their action decisions. All GenAI-assisted content, prompts, model outputs, and reported results were manually reviewed and verified by the authors, who remain responsible for the final paper.

\begin{acks}
This study used data recorded in a laboratory environment and did not involve any third-party participants. The data collection protocol was reviewed by the Institutional Review Board (IRB) of the corresponding author's institution (Project No. 20230024). To protect identifiable laboratory interiors and voices from the research team, we do not publicly release the raw audio or video; instead, we release the scenario manifest and model outputs to support reproduction. Raw media may be shared with researchers upon reasonable request.
\end{acks}

\balance
\bibliographystyle{ACM-Reference-Format}
\bibliography{main}

\appendix

\end{document}